\documentclass[a4paper]{jpconf}
\usepackage{graphicx}
\begin{document}
\title{Physics with prompt photons at SPD}

\author{Alexey Guskov on behalf of the SPD working group}

\address{Dzhelepov Laboratory of Nuclear Problems, Joint Institute for Nuclear Research, Joliot-Curie 6, Dubna Russia}

\ead{avg@jinr.ru}

\begin{abstract}
Prompt photons is a proven instrument for exploration of the gluonic structure of hadrons. The SPD experiment at the NICA collider plans to use prompt photons for study of the gluon contribution to the spin structure of the nucleon in the polarised proton-proton and deutron-deutron collisions. 
\end{abstract}

\section{Introduction}
NICA (Nuclotron-based Ion Collider fAсility) is a new accelerator complex designed at the Joint Institute
 for Nuclear Research (Dubna, Russia) to study properties of baryonic matter. In the first interaction point of the new collider 
 the MultiPurpose Detector  (MPD) 
 intends to study properties of hot dense nuclear matter
  in heavy ions collisions. The Spin Physics Detector (SPD) in the second interaction point will be constructed for
  investigation of the nucleon spin structure in collisions of longitudinally and transversely polarised protons and deuterons
   at $\sqrt{s}$ up to 27 GeV and luminosity up to 10$^{32}$ $s^{-1}cm^{-2}$. The measurement of 
   TMD PDFs for  quarks and gluons using such reactions as the Drell-Yan process, charmonia and prompt photon
   production is the main goal of the experiment \cite{loi,cdr}.  
   
   Prompt-photon production is a proven instrument for exploration of the gluonic structure of hadrons. Unpolarised and polarised physics with prompt photons and experimental conditions  of their registration in the SPD setup will be discussed below.

\section{Prompt photons at low energies}
Prompt photons are  photons produced in the hard scattering of partons. According to the factorization theorem, the inclusive cross section for the production of a prompt photon in a collision of hadrons $h_A$ and $h_B$ can be written as follows:
\begin{equation}
d\sigma_{AB\to \gamma X}=
\sum_{a,b=q,\bar{q},g}\int{dx_a dx_b}f_a^A(x_a,Q^2)f_b^B(x_b,\mu^2)d\sigma_{ab\to \gamma x}(x_a, x_b, Q^2).
\end{equation}
     The function $f_a^A$ ($f_b^B$) is the parton density for hadron $h_A$ ($h_B$), $x_a$ ($x_b$) is the fraction of the momentum of hadron $h_A$ ($h_B$) carried by parton $a$ ($b$) and $Q^2$ is  the square of the 4-momentum transferred in the hard scattering process, and $\sigma_{ab\to \gamma x}(x_a, x_b, Q^2)$ represents the cross section for the hard scattering of partons $a$ and $b$.  Prompt photon production in hadron collisions is one of the most direct ways to access the gluon structure of hadrons. 

Two main hard processes are responsible in the LO for the production of prompt photons in nuclear collisions: i) gluon Compton scattering, $gq(\bar{q})\rightarrow \gamma q(\bar{q})$, which dominates, and ii) quark-antiquark annihilation, $q \bar{q} \rightarrow \gamma g$. Contribution of the latter process to the total cross section does not exceed 20\% at the center-of-mass energy $\sim 20$ GeV/c$^2$. Hard processes with two prompt photons in the final state, such as $gg\to\gamma\gamma$ (via a box diagram) and  $q \bar{q} \rightarrow \gamma g$ (directly) contribute on a percent level. 

\begin{figure}[]
\begin{center}
\includegraphics[width=0.8\linewidth]{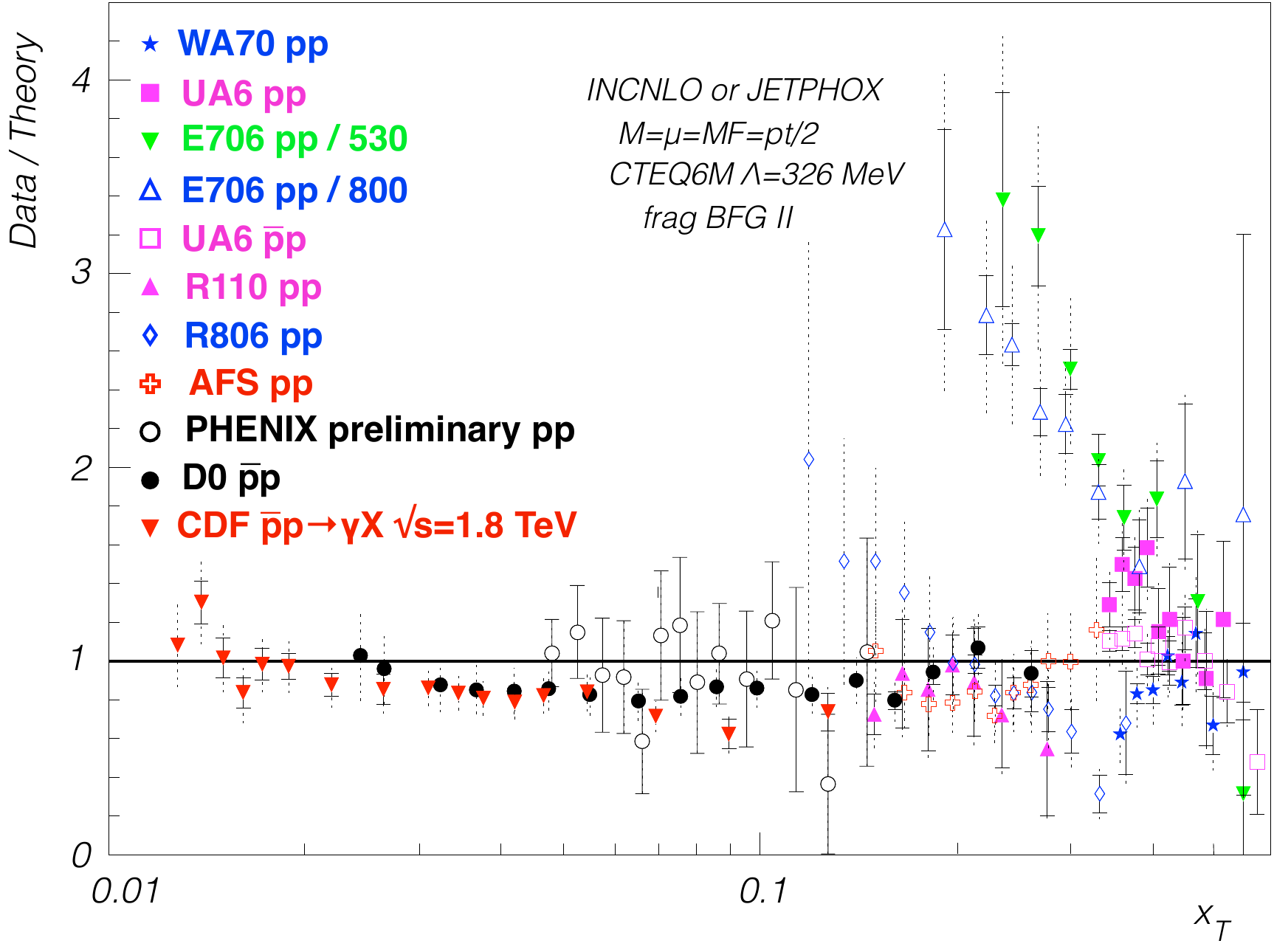}
\caption{Measured cross sections of prompt-photon production divided to the predicted those by theory as function of  $x_T$ \cite{Aurenche:2006vj}.}
\label{fig:pp_prompt_unpolarized}
\end{center}
\end{figure}

Unpolarised measurements of the differential cross section of prompt-photon production 
in proton-proton(antiproton) collisions were already performed by the fixed-target and collider experiments \cite{Vogelsang:1997cq}. Figure \ref{fig:pp_prompt_unpolarized} shows the ratio of the measured cross sections to the predicted by theory as function of  $x_T={2p_T}/\sqrt{s}$ \cite{Aurenche:2006vj} where $p_T$ is the transverse momentum of produced prompt photon. One can see that for fixed-target results corresponding to $\sqrt{s}\sim 20$ GeV, there is a significant disagreement with theoretical expectations that is absent for high-energy collider results. A new precise measurement at low energies could clarify the problem.

The measurement of single transverse spin asymmetry $A^{\gamma}_N$ in prompt-photon production at high $p_T$ in polarised $p$-$p$ and $d$-$d$ collisions could provide information on the gluon Sivers function, which is almost unknown at the moment \cite{Boer:2015vso}. The numerator of $A^{\gamma}_N$ can be expressed as \cite{Schmidt:2005gv}
\begin{eqnarray}
\nonumber
\sigma^{\uparrow}-\sigma^{\downarrow} &=&\sum \limits_{i}\int_{x_{min}}^1  dx_a \int
d^2{\bf k}_{Ta} d^2{\bf k}_{Tb}
\frac{x_a x_b}{x_a - (p_T/\sqrt{s})~e^y}
\left [ q_i(x_a,{\bf k}_{Ta}) \Delta_N G(x_b,{\bf k}_{Tb}) \right. \\
& & \times \frac{d\hat{\sigma}}{d \hat{t}}
 (q_i G \to q_i \gamma )
 +  \left.  G(x_a,{\bf k}_{Ta}) \Delta_N q_i (x_b,{\bf k}_{Tb})
 \frac{d\hat{\sigma}}{d \hat{t}} ( G q_i \to q_i \gamma  ) \right ]~~.
\end{eqnarray}
Here $\sigma^{\uparrow}$ and $\sigma^{\downarrow}$ are the cross sections of the prompt-photon production for the opposite transverse polarisations of one of the colliding protons,  $q_i(x,{\bf k}_{Ta})$ $[G(x,{\bf k}_{Ta})]$ is the quark [gluon] distribution function with specified ${\bf k}_{T}$ and $\Delta_N G(x_b,{\bf k}_{Tb})$ $[\Delta_N q_i (x_b,{\bf k}_{Tb})]$ is the gluon [quark] Sivers function. $d\hat{\sigma}/d \hat{t}$ represents the corresponding gluon Compton scattering cross section. The authors in \cite{Qiu:1991wg} pointed out that the asymmetry $A^{\gamma}_N$ at large positive $x_F$ is dominated by quark-gluon correlations while at large negative $x_F$ it is dominated by pure gluon-gluon correlations as it was concluded in \cite{Ji:1992eu}. The further development of the corresponding formalism can be found in \cite{Hammon:1998gb,Gamberg:2012iq}. It is important to notice that the corresponding known non-zero asymmetry in $\pi^0$ production, $A^{\pi^0}_N$ \cite{Adams1},  could be controlled in parallel.

The first attempt to measure $A^{\gamma}_N$ at $\sqrt{s}=19.4$ GeV was performed in the fixed target experiment E704 at Fermilab in the kinematic range $-0.15<x_F<0.15$ and 2.5 GeV/$c$ $<p_T<$ 3.1 GeV/$c$. The results were consistent with zero asymmetry within large statistical and systematic uncertainties \cite{Adams:1995gg}. 

\begin{figure}[h]
  \begin{minipage}[ht]{0.49\linewidth}
    \center{\includegraphics[width=1\linewidth]{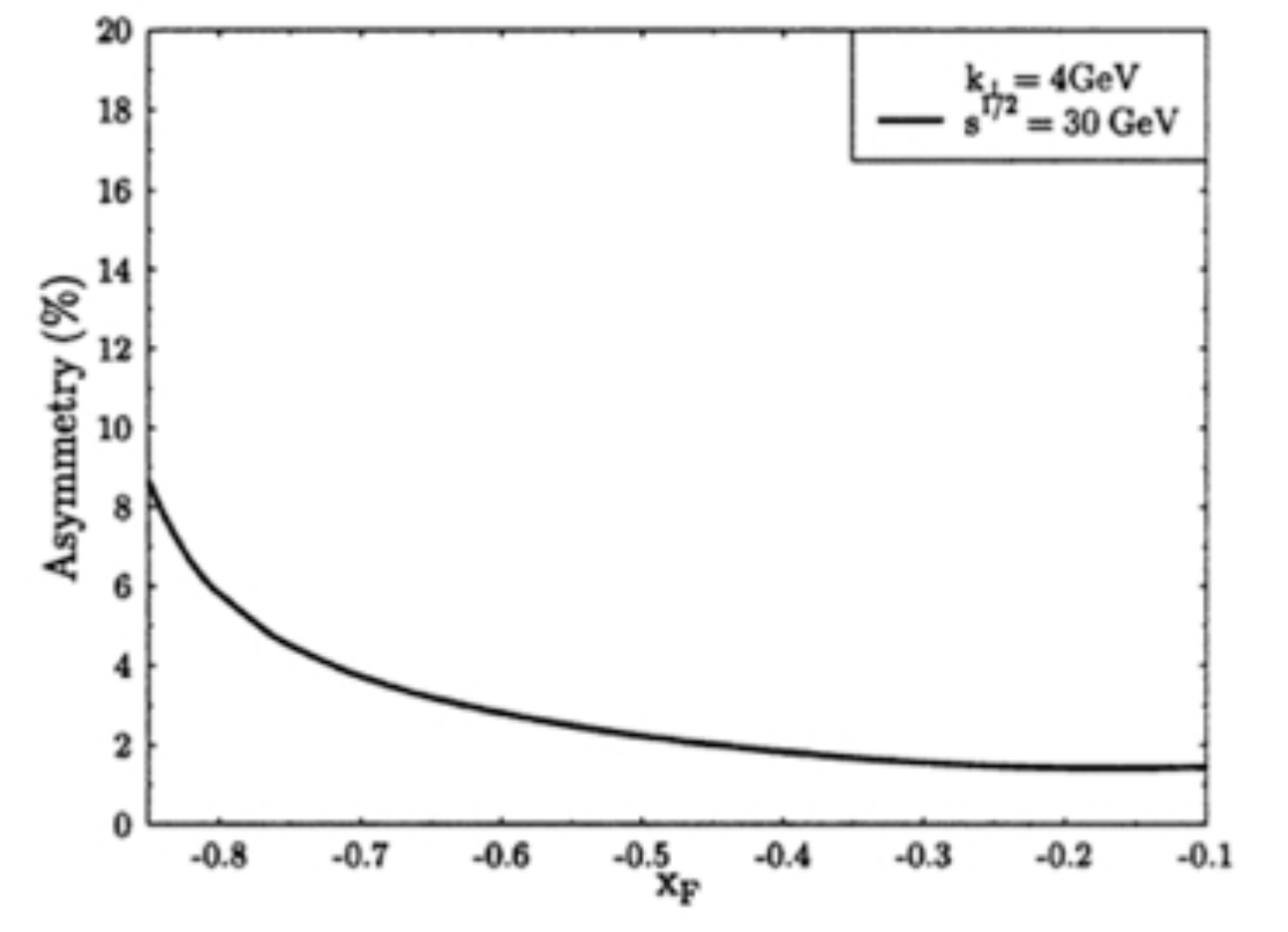} \\ (a)}
  \end{minipage}
  \hfill
  \begin{minipage}[ht]{0.49\linewidth}
    \center{\includegraphics[width=1\linewidth]{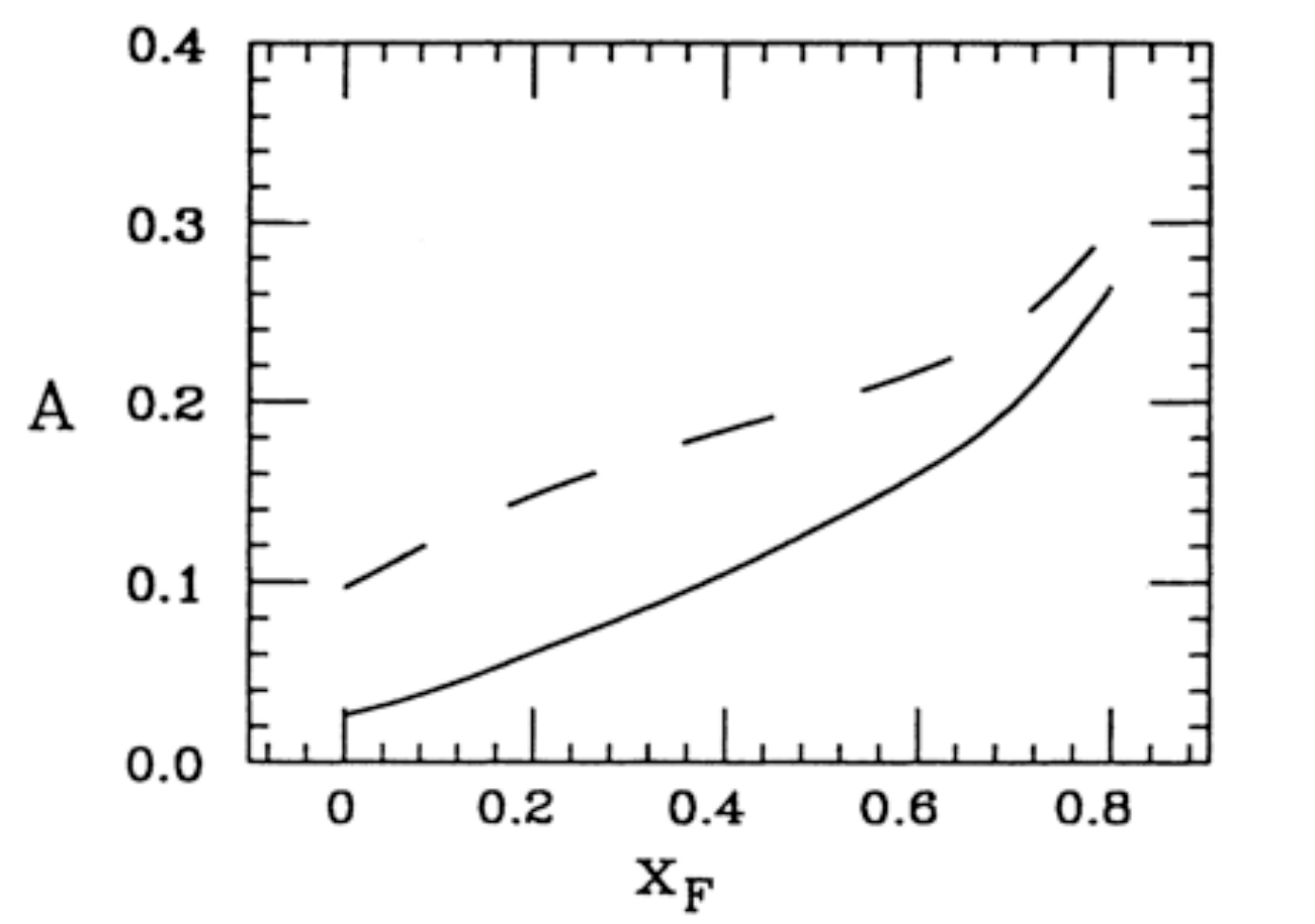} \\ (b)}
  \end{minipage}
  \caption{Theoretical predictions for $A^{\gamma}_N$ at $\sqrt{s}=30$ GeV and $p_T=4$ GeV/$c$ for (a) positive \cite{Hammon:1998gb} and (b) negative \cite{Qiu:1991wg} values of $x_F$.}
  \label{fig:pp_AN_theory}  
\end{figure}

     The study of prompt-photon production at large transverse momentum with longitudinally polarised proton beams could provide the access to gluon polarisation $\Delta{g}$ via the measurement of longitudinal double spin asymmetry $A^{\gamma}_{LL}$ \cite{Cheng:1989hf}. Assuming dominance of the gluon Compton scattering process, the asymmetry $A_{LL}^{\gamma}$ can be presented as \cite{Bunce:2000uv}

\begin{equation} \label{eqqg}
A_{LL} \approx \frac{\Delta g(x_a)}{g(x_a)} \, \cdot \, 
\Bigg[\frac{\sum_q e_q^2 \left[ \Delta q (x_b) + 
\Delta \bar{q} (x_b) \right]}{\sum_q e_q^2 \left[q (x_b)+
\bar{q} (x_b) \right] } \Bigg] \, \cdot \, 
\hat{a}_{LL}(gq \rightarrow \gamma q) + (a\leftrightarrow b) \, .
\end{equation} 
The second factor in the equation coincides to the lowest 
order with the spin asymmetry $A_1^p$ well-known from polarised DIS, 
and the partonic asymmetry $\hat{a}_{LL}$ is calculable in the perturbative QCD. Previous results for gluon polarisation show that gluon polarisation is consistent with zero: $|\Delta{g}/g|<\pm0.2$ while the asymmetry $A_1^p$ is about 0.2. So it seems that the value of $A^{\gamma}_{LL}$ should not exceed the level of a few percent. Measurement of $A_{LL}^{\gamma}$ asymmetry as well as $A^{\gamma}_N$ is a part of the spin physics programme at the RHIC collider \cite{JalilianMarian:2000qa,Sakashita:2009nba,Horaguchi:2006zz,Skoro:1999rg,Bunce:2000uv}.

Authors in \cite{Li:2008ym,Lansberg:2015hla} propose also to pay attention to associative prompt-photon production in reactions like $p p \to \gamma J/\psi X$.

\section{Prompt photons at SPD}
The SPD setup is planned as a universal 4$\pi$ detector.
It consists of three parts: barrel and two end-caps. Each part has an individual magnetic system. 
The toroidal field should provide tracking capability in the barrel part while the end-caps are equipped by
solenoidal coils. Such configuration of the magnetic system should minimize the magnetic field in the beam interaction region. Tracking system (TS) of the SPD setup consists of the silicon-based inner tracker (IT) in the central part of the detector surrounded by the main tracker using gas-filled drift straw-tubes as the basic detection element. The time-of-flight system (TOF) should provide identification of secondary hadrons in a wide kinematic range. The shashlyk-type electromagnetic calorimeter (ECAL) is responsible for photon reconstruction. It will be the main element for the described programme with prompt phonons.  The range (muon) system (RS) should perform the advanced muon identification via muonic pattern recognition and further matching of track segments to tracks in the inner part of the detector. 

\begin{figure}[!t]
  \begin{minipage}[ht]{0.49\linewidth}
    \center{\includegraphics[width=1\linewidth]{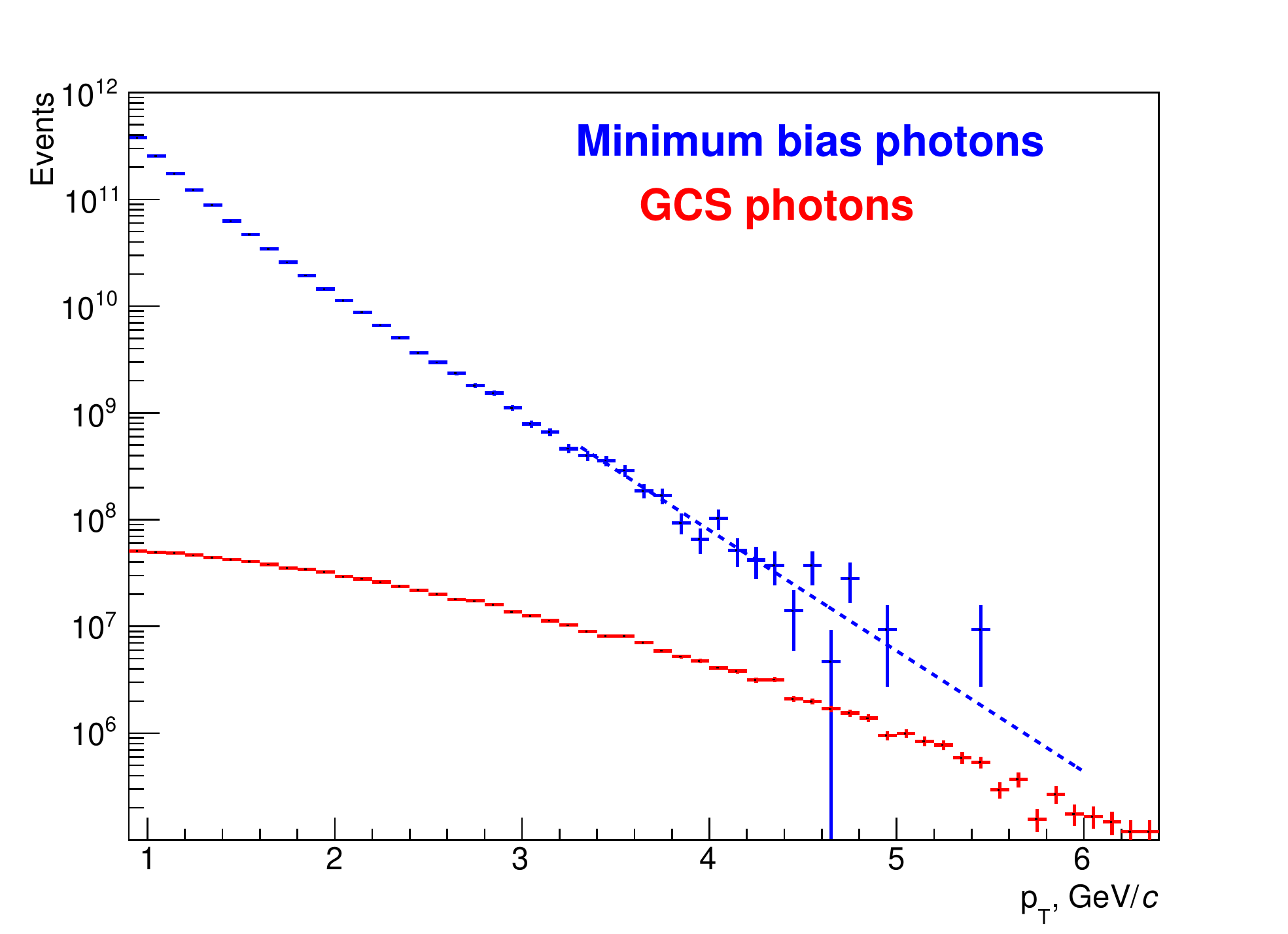} \\ (a)}
  \end{minipage}
  \hfill
  \begin{minipage}[ht]{0.49\linewidth}
    \center{\includegraphics[width=1\linewidth]{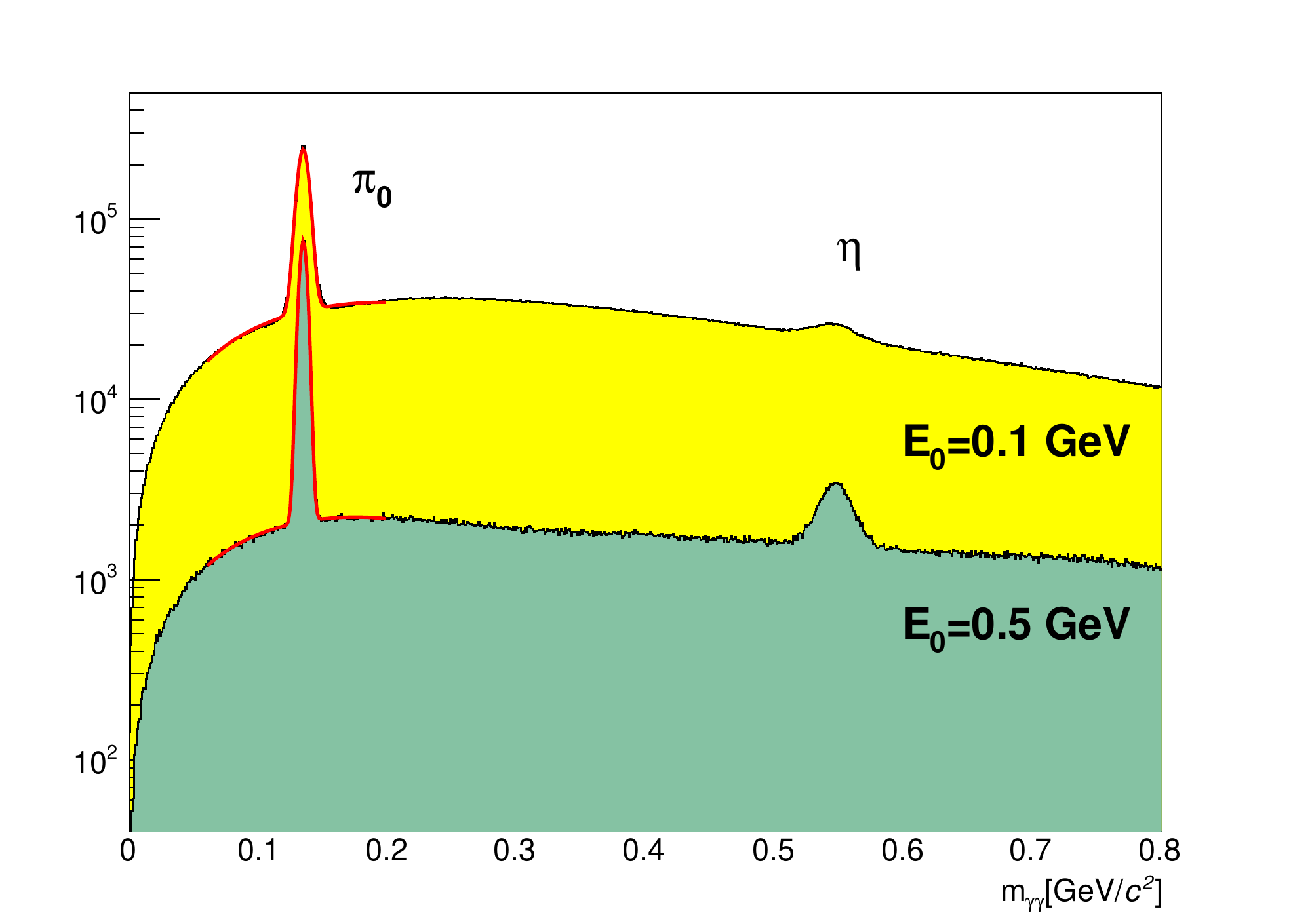} \\ (b)}
  \end{minipage}
  \caption{ (a) The $p_T$ spectra of GCS prompt photons (red) and photons produced from $\pi^0$ decay (blue). (b)  
  The invariant mass spectrum of two photons for two ECAL thresholds: 100 MeV (yellow) and 500 MeV (green).
}
  \label{fig:pp_rec1}  
\end{figure}

The gluon Compton scattering (GCS) is the main mechanism of the prompt-photon production at SPD energies. At $\sqrt{s}=26$ GeV, the corresponding cross section calculated for $p_T>1$ GeV/$c$ in the leading order is 1.2 $\mu$b. The contribution of the competitive quark-antiquark annihilation process is one order of magnitude smaller. Nevertheless the main source of photons in hadronic collisions is the decay of secondary particles, mainly $2\gamma$ decay of $\pi^0$ and $\eta$ mesons. Contributions of  photons from decays of  different secondary particles  per one interaction are presented in Tab.~\ref{tab:gamma_bkg}. The $p_T$ spectra of the GCS photons and the decay photons produced near the interaction point are shown in Fig. \ref{fig:pp_rec1}(a). The contribution of the decay photons dominates over the GCS signal in full range of $p_T$ but relative fraction of signal photons  increases with increasing $p_T$. The rejection of photons from reconstructed 2$\gamma$ decays of $\pi^0$ and $\eta$-mesons and the following Monte Carlo-based statistical subtraction of residual background photons is the only a way to access the prompt-photon production cross section at high $p_T$.
\begin{table}[htp]
\caption{Contributions of  photons from decays of  different secondary particles per one inelastic interaction (according to Pythia6). }
\begin{center}
\begin{tabular}{|c|c|}
\hline
Decay  & Number of photons \\
\hline
$\pi^0$ & 8.70 \\
$\eta$  & 0.40 \\
$\omega$ & 0.06 \\
$\eta'$  &  0.05 \\
$\Sigma^0$ & 0.04 \\
Others    &  0.01 \\
\hline
\end{tabular}
\end{center}
\label{tab:gamma_bkg}
\end{table}%

Other important sources of background clusters in the electromagnetic calorimeter are:
\begin{itemize}
\item clusters produced in the interaction of secondary particles with the material of the setup elements;
\item ''charged'' clusters misidentified as neutral in the ECAL due to inefficiency of track finding and reconstruction algorithms;
\item overlapping of clusters produced by different particles;
\item clusters produced by neutral hadrons.
\end{itemize}

For successful implementation of the prompt-photon programme the SPD setup and the ECAL have to meet the following requirements:
\begin{itemize}
\item inner space  of the SPD setup should contain as minimal amount of material as possible in order to keep high transparency of the setup for photons;
\item minimal tracking capability should be provided in order to distinguish charged and neutral clusters in ECAL;
\item reasonable granularity of the ECAL in order to avoid the pile-up effect that reduces the photon reconstruction efficiency 
and distorts kinematics of reconstructed photons;
\item the ability to  detect  with reasonable resolution an energy deposit down to 100 MeV to keep high reconstruction efficiency for $\pi^0$ decays.  It is important to note that this threshold is significantly below the MIP signal ($\sim$ 200~MeV) for the planned calorimeter. 
\end{itemize}

\begin{figure}[]
  \begin{minipage}[ht]{0.49\linewidth}
    \center{\includegraphics[width=1\linewidth]{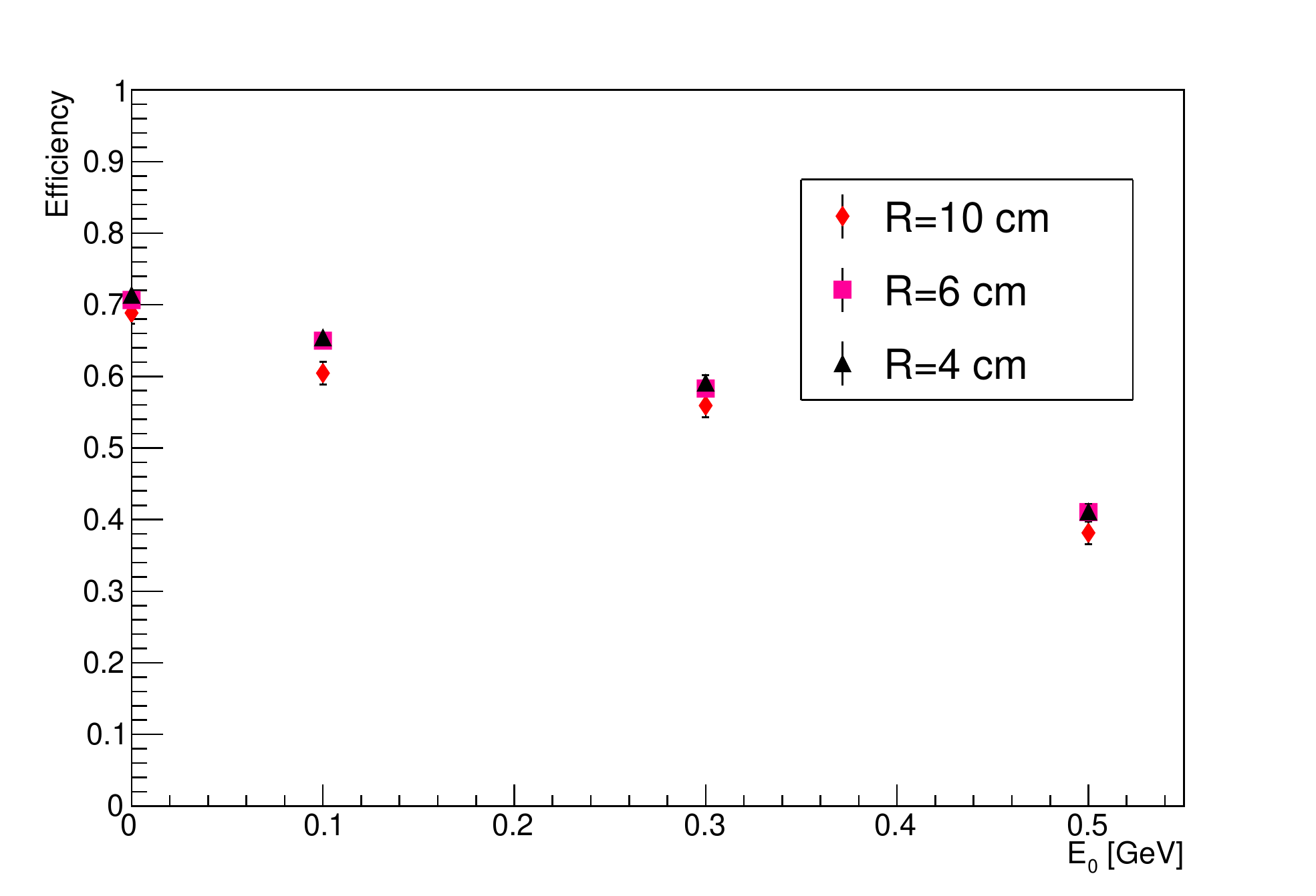} \\ (a)}
  \end{minipage}
  \hfill
  \begin{minipage}[ht]{0.49\linewidth}
    \center{\includegraphics[width=1\linewidth]{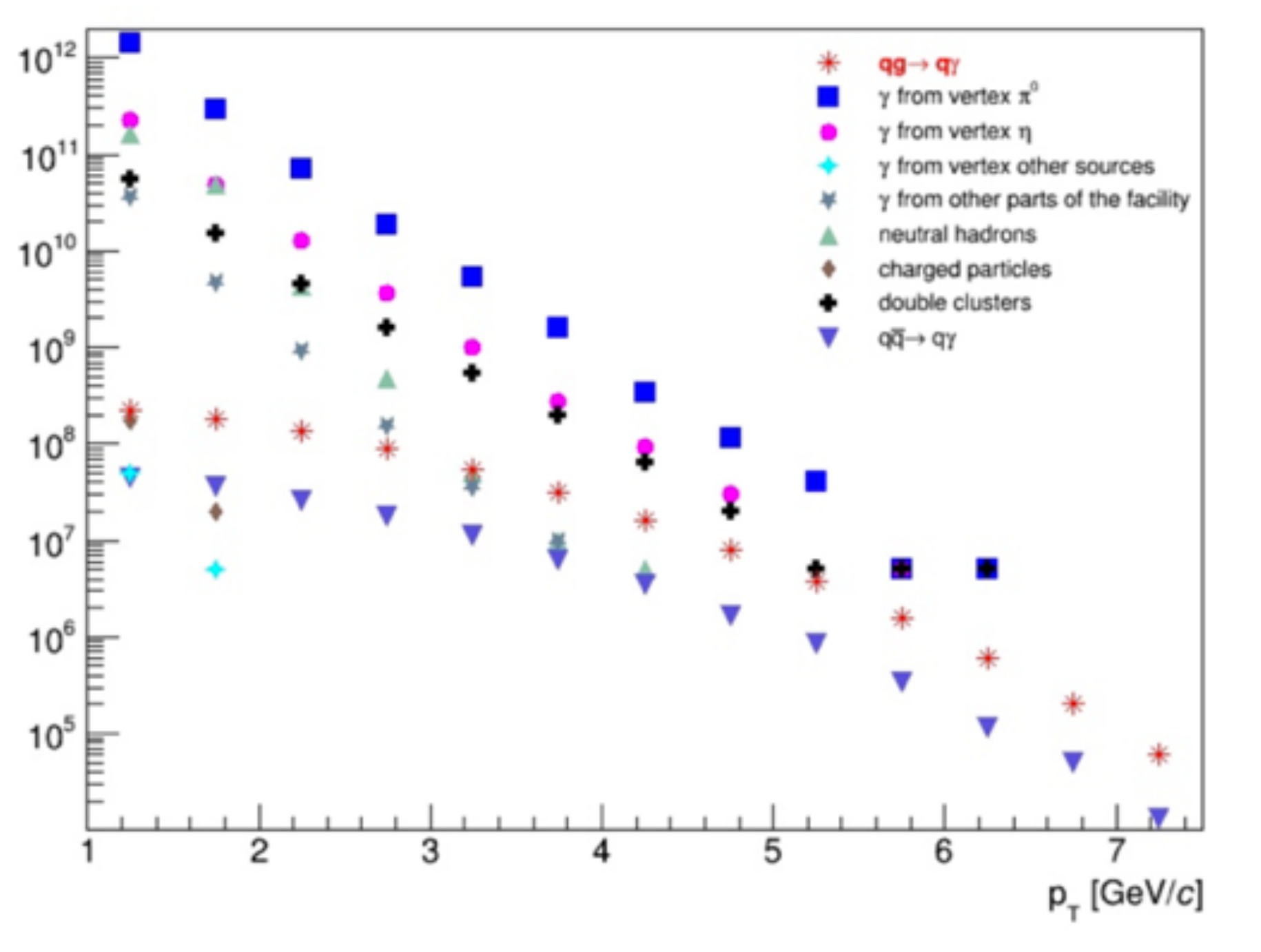} \\ (b)}
  \end{minipage}
  \caption{ (a) The efficiency of the $\pi^0\to\gamma\gamma$ decay reconstruction in case when at least one photon has $p_T>2$ GeV/$c$ as a function of cluster energy threshold for different granularities R of the ECAL.   (b)  }
  \label{fig:pp_rec2}  
\end{figure}

The invariant mass spectrum of two photons reconstructed in the ECAL with 100 MeV and 200 MeV  energy threshold is shown in Fig.~\ref{fig:pp_rec1}(b). Relative energy resolution of the ECAL was assumed to be $\sigma_E/E=1.96\% \oplus 2.74\%/\sqrt{E/GeV}$. Gaussian width of the $\pi^0$ peak is 4.6 MeV/$c^2$ and 3.15 MeV/$c^2$ for 100 MeV and 500 MeV thresholds, respectively. Figure \ref{fig:pp_rec2}(a) shows the efficiency of the $\pi^0\to\gamma\gamma$ decay reconstruction for different cluster energy thresholds and different granularities of the ECAL. More kinematic distributions could be found at \cite{Guskov13}. Contributions of the signal and each kind of background mentioned above are shown in  Fig. \ref{fig:pp_rec2}(b). Tracking efficiency is assumed to be on the level of 90\% while the ECAL cell size is 4 cm \cite{Rymbekova}.

The accuracy of the prompt-photon production cross section and asymmetries $A_N$ and $A_{LL}$ in $p$-$p$ collisions is estimated on the basis of the following assumptions. Data sample corresponds to $10^{7}$ s of data taking (about 100 days) with average luminosity $L=10^{32}~s^{-1}cm^{-2}$. The inefficiency of the photon reconstruction is caused only by geometrical acceptance and material distribution, no threshold for photon reconstruction in the ECAL. LO calculations are used for the GCS cross-section. Only $\pi^0\to \gamma\gamma$ decay is taken into account as background. Average $\pi^0$ reconstruction efficiency is about 70\%. Relative accuracy for the Monte Carlo description of photon reconstruction is assumed to be 2\%. Two subsamples with equal statistics are collected with each combination of spin orientations.  Absolute polarisation of the beam is assumed to be exactly equal to 100\%. We also ignore any uncertainties related with the control of the beam polarisation(s) and luminosity.

Since low-$p_T$ region is useless due to a huge background while at high-$p_T$ the statistics is very limited, a reasonable cut on transverse momentum of photon has to be applied in order to maximize the accuracy of the planned measurements. For the running conditions mentioned above the optimal value lies in the range from 5 to 6 GeV/$c$. The total amount of the detected GCS photons with $p_T>5$ GeV/$c$ is expected to be about $6\times10^6$. 

\begin{figure}[]
  \begin{minipage}[ht]{0.49\linewidth}
   \center{\includegraphics[width=1\linewidth]{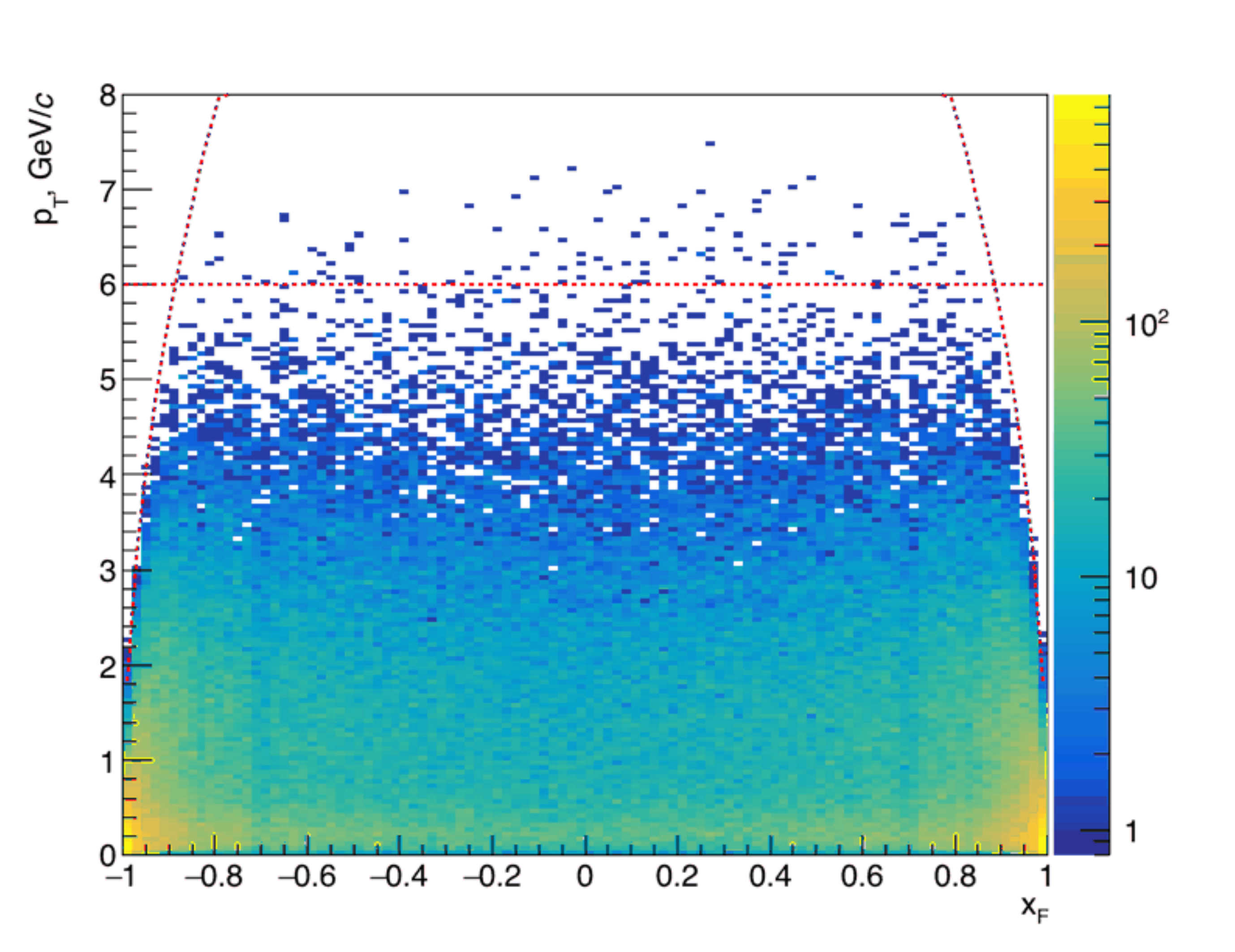}  \\ (a)}
   \end{minipage}
   \hfill  
     \begin{minipage}[ht]{0.49\linewidth}
   \center{\includegraphics[width=1\linewidth]{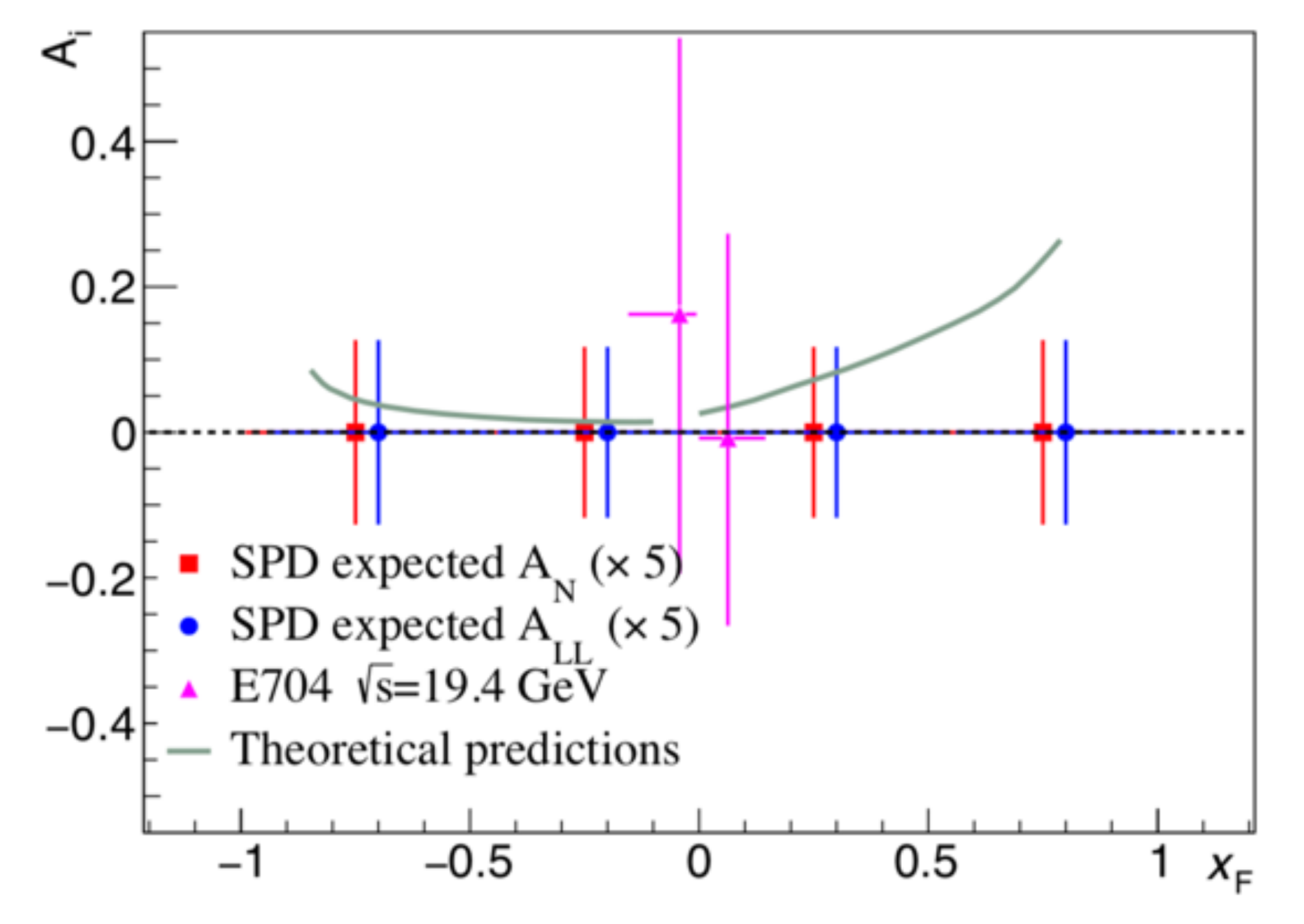}  \\ (b)}
   \end{minipage}
  \caption{ (a) $p_T$ distribution for GCS photons vs. $x_F$. (b) Expected accuracy of $A_{N}$ and $A_{LL}$ measurement as a function of $x_{F}$.}
  \label{fig:pp_rec3}  
\end{figure}

Figure \ref{fig:pp_rec3}(a) shows the distribution of $p_T$ vs. $x_F$ for GCS photons.The expected accuracy of $A_N$ and $A_{LL}$ measurement in each of four intervals in $x_F$ in the range from -0.89 to 0.89 is shown in Fig \ref{fig:pp_rec3}(b). The results of previous measurement and the theoretical expectations are also presented on the plot.   

\section{Summary}
 Unpolarised and polarised physics with prompt photons looks very attractive since 
 all previous the measurements at energy scale of about 20 GeV were performed 20--30 years ago. It is a good time to come back with a new level of experimental techniques and theoretical understanding. The SPD detector at the NICA collider provides a good chance to perform such kind of measurements. Preliminary analysis of the background conditions shows that the measurement of asymmetries in the prompt-photon production cross section on the level of a few per cent is possible at SPD conditions. Taking into account quite minimalistic requirements to the tracking system of the SPD detector, the study of the prompt-photon production could be the first stage of the SPD setup operation.


\section*{References}
\begin{thebibliography}{9}
\bibitem{loi} 
Savin I, Efremov A, Peshekhonov D, Kovalenko A, Teryaev O, Shevchenko O, Nagajcev A, Guskov A, Kukhtin~A and Toplilin N 2015
{\it EPJ Web of Conferences} {\bf 85} 02039
\bibitem{cdr}
SPD working group 2018
{\it Conceptual design of the Spin Physics Detector} 
\bibitem{Vogelsang:1997cq}
  Vogelsang W and Whalley M R 1997
 {\it J.\ Phys.\ } G {\bf 23}  A1
\bibitem{Aurenche:2006vj}
  Aurenche P, Fontannaz M, Guillet G P, Pilon E and Werlen M 2006
  {\it Phys.  Rev.}~D {\bf 73}  094007
\bibitem{Boer:2015vso} 
  Boer D, Lorcé C, Pisano C and Zhou J 2015 
 {\it Adv.\ High Energy Phys.}  {\bf 2015}, 371396
 \bibitem{Schmidt:2005gv}
  Schmidt I, Soffer J and Yang J J 2005
  {\it Phys.\ Lett.\ }~B {\bf 612}  258
\bibitem{Qiu:1991wg}
  Qiu J W and Sterman G F 1992
  {\it Nucl.\ Phys.\ }~B {\bf 378} 52
  \bibitem{Ji:1992eu}
  Ji X D 1992
  {\it Phys.\ Lett.\ }~B {\bf 289} 137
 \bibitem{Hammon:1998gb}
  Hammon N, Ehrnsperger B and Schaefer A 1998
  {\it J.\ Phys.\ }~G {\bf 24}  991
\bibitem{Gamberg:2012iq}
  Gamberg L and Kang Z B 2012
  {\it Phys.\ Lett.\ } B {\bf 718} 181
 \bibitem{Adams1}
  Adams D L {\it et al.} [E581 and E704 Collaborations] 1991
  {\it Phys.\ Lett.\ } B {\bf 261} (1991) 201
 \bibitem{Adams:1995gg}
  Adams L G {\it et al.} [E704 Collaboration] 1995
  {\it Phys.\ Lett.\ }B {\bf 345} 569
\bibitem{Cheng:1989hf}
  Cheng H Y and Lai S N 1990
  {\it Phys.\ Rev.\ }D {\bf 41}  91  
\bibitem{Bunce:2000uv}
  Bunce G, Saito N, Soffer J and Vogelsang W 2000
  {\it Ann.\ Rev.\ Nucl.\ Part.\ Sci.\  } {\bf 50} 525
\bibitem{JalilianMarian:2000qa}
  Jalilian-Marian J, Orginos K and Sarcevic I 2001
  {\it Phys.\ Rev.\ } C {\bf 63} 041901
\bibitem{Sakashita:2009nba}
  Sakashita K 2009 
  {\it Cross Section for Prompt Photon Production in Proton-Proton Collisions at $\sqrt{s}$ = 62.4~GeV} PhD thesis
\bibitem{Horaguchi:2006zz}
  Horaguchi T 2006
  {\it Prompt photon production in proton proton collisions at $\sqrt{s}$ = 200~GeV} PhD thesis
\bibitem{Skoro:1999rg}
  Skoro G P, Zupan M and Tokarev M V 1999
  {\it Nuovo Cim.\ } A {\bf 112}  809
\bibitem{Li:2008ym}
  Li R and Wang J X 2009 {\it Phys.\ Lett.\ } B {\bf 672} 51
\bibitem{Lansberg:2015hla}
  Lansberg J P 2016 
 {\it Int.\ J.\ Mod.\ Phys.\ Conf.\ Ser.\  }{\bf 40} 1660015
 \bibitem{Guskov13}
Guskov A 2014 {\it Proc. XV Advanced Research Workshop on High Energy Spin Physics
DSPIN-13)} p 363
 \bibitem{Rymbekova}
Rymbekova A 2019 {\it Ukr. J. Phys.} {\bf 64} 631

  \end{thebibliography}
\end{document}